\documentclass[showpacs,preprintnumbers,amsmath,amssymb,twocolumn]{revtex4}
\usepackage{amsfonts}

\usepackage{graphicx}
\usepackage{dcolumn}
\usepackage{bm}
\usepackage{amssymb}
\usepackage{array}
\usepackage{longtable}

\begin{document}

\preprint{APS/123-QED}

\title{Counting spanning trees in self-similar networks by evaluating determinants}

\author{Yuan Lin}

\author{Bin Wu}

\author{Zhongzhi Zhang}
\email{zhangzz@fudan.edu.cn}
\homepage{http://homepage.fudan.edu.cn/~zhangzz/}

\affiliation {School of Computer Science, Fudan University, Shanghai
200433, China}

\affiliation {Shanghai Key Lab of Intelligent Information
Processing, Fudan University, Shanghai 200433, China}

\author{Guanrong Chen}
\email{eegchen@cityu.edu.hk}
\affiliation {Department of Electronic Engineering, City University of Hong Kong, Hong Kong, China}




\date{\today}

\begin{abstract}
Spanning trees are relevant to various aspects of networks. Generally, the number of spanning trees in a network can be obtained by computing a related determinant of the Laplacian matrix of the network. However, for a large generic network, evaluating the relevant determinant is computationally intractable. In this paper, we develop a fairly generic technique for computing determinants corresponding to self-similar networks, thereby providing a method to determine the numbers of spanning trees in networks exhibiting self-similarity. We describe the computation process with a family of networks, called $(x,y)$-flowers, which display rich behavior as observed in a large variety of real systems. The enumeration of spanning trees is based on the relationship between the determinants of submatrices of the Laplacian matrix corresponding to the $(x,y)$-flowers at different generations and is devoid of the direct laborious computation of determinants. Using the proposed method, we derive analytically the exact number of spanning trees in the $(x,y)$-flowers, on the basis of which we also obtain the entropies of the spanning trees in these networks. Moreover, to illustrate the universality of our technique, we apply it to some other self-similar networks with distinct degree distributions, and obtain explicit solutions to the numbers of spanning trees and their entropies. Finally, we compare our results for networks with the same average degree but different structural properties, such as degree distribution and fractal dimension, and uncover the effect of these topological features on the number of spanning trees.




\end{abstract}

\maketitle


\section{introduction}

Counting spanning trees in networks is a fundamental issue in both theory and applications, which has recently attracted increasing attention from mathematics~\cite{BuPe93,BrMa97,Ly05,LyPeSc08}, physics~\cite{Wu77,ShWu00,ChSh03,ChChYa07,TeWa10,TeWa11}, and other fields~\cite{JaTh89}. As a network invariant, spanning tree is a crucial measure of the network reliability~\cite{Bo86,SzAlKe03,BoSaSu09}. The notion is also closely related to various aspects of networks. For example, the number of spanning trees of a connected network is exactly the number of recurrent configurations of the Abelian sandpile model on the network~\cite{Dh90,Dh06}, which is equivalent to the chip-firing game~\cite{Me05} under certain restrictions and is a paradigm for self-organized criticality~\cite{BaTaWi87}. As another example, spanning trees offer useful insights into understanding the origin and mechanism of fractality in fractal scale-free networks~\cite{GoSaKaKi06}. In addition, spanning trees are relevant to other interesting problems on networks, such as transport~\cite{WuBrHaSt06}, electrical networks~\cite{BaGu10}, unbiased random walks~\cite{NoRi04,CoBeTeVoKl07}, loop-erased random walks~\cite{DhDh97}, and so on.

In view of their relevance to diverse aspects of networks and a wide range of applications~\cite{WuCh04}, spanning trees in networks  have become a focus of some recent research~\cite{DoDu01,Wu02,KiNoJe04,MaAlBa05,SeBoVe09, Kr10, OzYa11}. Particularly, in the physics literature a lot of efforts have been devoted to enumerating spanning trees in specific networks by using different techniques according to their special structures. Examples include regular lattices~\cite{Wu77,ShWu00,ChSh03,TeWa10}, Sierpinski gaskets~\cite{ChChYa07,TeWa11}, Erds\"o-R\'enyi random graphs~\cite{LyPeSc08}, the pseudofractal
scale-free web~\cite{ZhLiWuZh10}, fractal scale-free networks~\cite{ZhLiWuZo11}, and so on. These works have provided some effective methods for determining spanning trees in special networks, and have revealed some nontrivial effects of network structural properties on spanning trees.

Most existing methods for counting the number of spanning trees are applicable only to particular networks. Due to the complexity and diversity of networks, a general approach for counting spanning trees of a generic network is not available, with the exception of the Kirchhoff Matrix Tree Theorem~\cite{Ki1847}.  This well-known theorem provides a universal algorithm for determining the number of spanning trees of an arbitrary connected graph, in terms of a determinant~\cite{ChKl78}. However, the computational complexity for evaluating the determinant of a general network is very high~\cite{Bi93}. For large networks, it is difficult and even impossible to obtain their numbers of spanning trees. For this reason, seeking an efficient and somewhat general technique for calculating the determinant of a network is a matter of exceptional importance.

On the other hand, it has been observed that by nature most real networks are fractal~\cite{SoHaMa05} as well as scale-free~\cite{BaAl99}. Both striking properties have significant influences on other structural properties, e.g., average distance~\cite{ChLu02,CoHa03,ZhZhZoChGu09} and degree correlations~\cite{SoHaMa06}.  In particular, existing results have shown that in some special networks, the scale-free property may decrease the number of spanning trees~\cite{ZhLiWuZh10}, while fractality can increase the number of spanning trees~\cite{ZhLiWuZo11}. Thus, some interesting questions rise naturally: Can the power-law behavior alone determine the number of spanning trees in a scale-free network? What is the relation between the fractal dimension and the spanning trees in a fractal scale-free network? Both questions remain unsolved today.

In this paper, we present an approach for explicitly determining the number of spanning trees by evaluating a related determinant in a network, which is expected to hold true for a broad class of self-similar graphs. Our method is based on the recursion relations of determinants of submatrices of the Laplacian matrix associated with the network at different iterations, which avoids the laborious computation of the largest determinant as needed by classic algebraic schemes. To the best of our knowledge, this constitutes the first work in exploring fast evaluation of the determinant of a complex network, especially for networks with scale-free~\cite{BaAl99}, small-world~\cite{WaSt98}, and fractal~\cite{SoHaMa05} properties.

To demonstrate the computation process of our technique, we apply it to find the number of spanning trees in the family of self-similar networks, called $(x,y)$-flowers~\cite{RoHaAv07,RoAv07}, which display some remarkable properties observed in real-life networks~\cite{AlBa02,DoMe02,Ne03} and include the extensively studied pseudofractal scale-free web~\cite{DoGoMe02,ZhZhCh07} and fractal hierarchical lattices~\cite{BeOs79,KaGr81,GrKa82,HiBe06} as their limiting cases.  Using the proposed technique, we derive the exact number of spanning trees in the whole family of $(x,y)$-flowers, based on which we obtain their entropies of spanning trees. In addition, in order to exhibit the generality of our method, we apply it to some other self-similar networks and obtain closed-form solutions for the numbers of spanning trees in such networks. Finally, we present a detailed analysis of the obtained results, and show that power-law distribution alone do not suffice to characterize the numbers of spanning trees in scale-free networks, and that the increasing fractal dimensions of fractal  scale-free networks may lead to decrements of the numbers of spanning trees.

\section{\label{SecModel}Model and properties of $(x,y)$-flowers}

\begin{figure}[h]
\includegraphics[width=0.45\linewidth,trim=100 0 100 10]{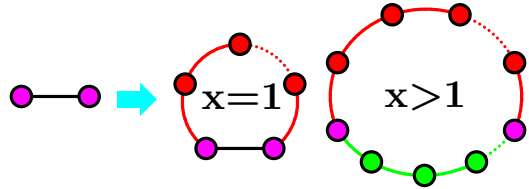}
\caption{(Color online) Iterative construction method of the $(x,y)$-flowers. Each edge is
replaced by two parallel paths with lengths $x$ ($x\geq 1$) and $y$ ($y\geq x$ and $y>
1$) on the right-hand side of the arrow.  For $x=1$, a pair of old nodes directly connected by an old edge generates $y-1$ new nodes. All of these $y-1$ new nodes and the two old nodes form a red path of length $y$; while the old edge is remained. For $x>1$, each old edge connecting two old nodes is removed and replaced by two paths with the two old nodes as the ends of the two paths: the $y-1$ red nodes and the two old nodes form a path of length $y$, while the $x-1$ green nodes and the two old nodes constitute another path of length $x$. }\label{fig1}
\end{figure}

\begin{figure}
\begin{center}
\includegraphics[width=0.60\linewidth,trim=0 0 0 0]{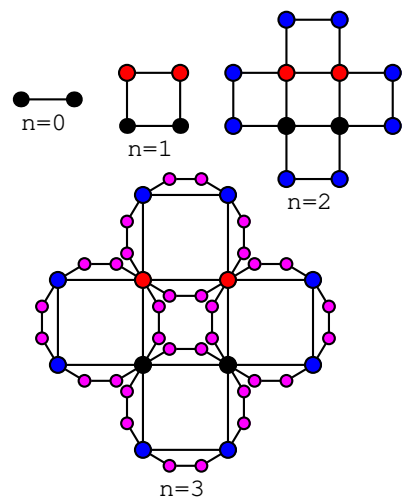}
\caption{(Color online) Illustration of the growing process for the $(1,3)$-flower.} \label{flower}
\end{center}
\end{figure}

\begin{figure}[h]
\centering
\includegraphics[width=0.8\linewidth,trim=0 0 0 0]{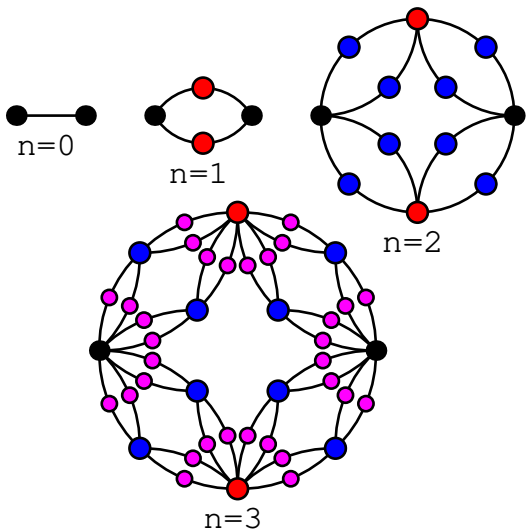}
\caption{(Color online) Sketch of the iterative process for the $(2,2)$-flower.}\label{fractal}
\end{figure}

In this section, we introduce the construction and structural properties of the $(x,y)$-flowers~\cite{RoHaAv07,RoAv07}, which are built in an iterative way. Let $F_{n}(x,y)$ ($n \geq 0$) denote the $(x,y)$-flowers after $n$ generations of evolution. In what follows it is assumed that $x \le y$ and $y>1$ without loss of generality. The $(x,y)$-flowers are constructed as follows. For $n=0$, $F_{0}(x,y)$ is an edge connecting two nodes, called the initial nodes hereafter. For $n \ge 1$, $F_{n}(x,y)$ is derived from $F_{n-1}(x,y)$ through replacing each exiting edge in $F_{n-1}(x,y)$ by two parallel paths of lengths $x$ and $y$, see Fig.~\ref{fig1}. For illustration, in Figs.~\ref{flower} and~\ref{fractal}, we present the growing processes of two particular networks:  $(1,3)$-flower and $(2,2)$-flower. According to the generating algorithm for the $(x,y)$-flowers, it is easy to see that the number of edges in $F_{n}(x,y)$ is
\begin{equation}\label{Mn}
M_{n} = (x+y)^n\,.
\end{equation}

\begin{figure}[h]
\centering
\includegraphics[width=0.9\linewidth,trim=0 0 0 0]{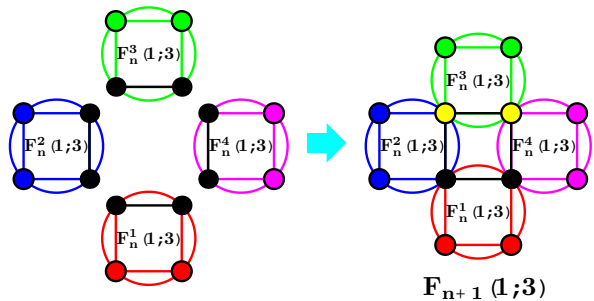}
\caption{(Color online) Illustration of the alternative construction algorithm for the $(1,3)$-flower.  $F_{n+1}(1,3)$ is obtained through amalgamating four copies of $F_{n}(1,3)$, denoted by $F_{n}^{\eta}(1,3)$ ($\eta=1,2,3,4$), by merging four pairs of the initial nodes belonging to different $F_{n}^{\eta}(1,3)$. Black nodes represent the initial nodes of $F_{n}^{\eta}(1,3)$ or of $F_{n+1}(1,3)$.}\label{Alternative}
\end{figure}

Alternatively, the $(x,y)$-flowers can be constructed by using the following equivalent algorithm~\cite{RoHaAv07}, which highlights the self-similarity of the networks. From $F_{n-1}(x,y)$, the next generation $F_{n}(x,y)$ is obtained by joining $x+y$ copies of $F_{n-1}(x,y)$ at the two initial nodes of each replica. 
Figure~\ref{Alternative} provides an illustration for the second construction approach for the $(1,3)$-flower. 
According to this construction, the number of nodes $N_n$ in $F_{n}(x,y)$ satisfies the recursive relation $N_n=(x+y)N_{n-1}-(x+y)$, which together with the initial condition $N_1=x+y$ yields
\begin{equation}\label{Nn}
N_{n}=\frac{x+y-2}{x+y-1}(x+y)^{n}+\frac{x+y}{x+y-1}\,.
\end{equation}
Thus, the average degree of the $(x,y)$-flowers is
\begin{equation}\label{kn}
\langle k \rangle _{n} = \frac{2M_{n}}{N_{n}}=\frac{2(x+y-1)(x+y)^{n-1}}{(x+y-2)(x+y)^{n-1}+1}\,,
\end{equation}
which approaches $2(x+y-1)/(x+y-2)$ as $n\rightarrow \infty$.

The deterministic construction also allows to determine the full
degree distribution of the $(x,y)$-flowers. In $F_{n}(x,y)$, the degree spectra of all nodes are discrete, i.e., nodes have only degree $k$ in the form of $k=2^m$ ($m=1,2,\ldots,n$). Let $N_n(m)$ be the number of nodes with degree $2^m$ in $F_{n}(x,y)$. Then, $N_n(m)=(x+y-2)(x+y)^{n-m}$ for $m <
n$, and $N_n(m)=x+y$ for $m=n$. Thus, in the family of $F_{n}(x,y)$, all networks with the same $x+y$ have an identical degree sequence, hence an identical degree distribution.

The $(x,y)$-flowers exhibit rich behavior in their architecture~\cite{RoHaAv07,RoAv07}. They obey a power-law degree distribution $P(k) \sim k^{-\gamma}$ with the exponent $\gamma=1 +\ln(x+y)/\ln2$ lying in the interval $[1+\ln 3/ \ln2,\infty)$. For $x=1$, the networks are small-world but non-fractal; while for $x>1$, they are ``large-world" and fractal with the fractal dimension $d_{\rm f}=\ln(x+y)/\ln x$. These topological properties are not shared by any other modeling networks, which make the $(x,y)$-flowers unique within the family of networks.

\section{Spanning trees in $(x,y)$-flowers}

After introducing the construction and some characteristics of the
$(x,y)$-flowers, we now attempt to determine analytically the number of spanning trees in
$F_{n}(x,y)$, denoted by $N_{\rm ST}(n)$. According to the Matrix Tree Theorem~\cite{Ki1847,ChKl78}, $N_{\rm ST}(n)$ can be evaluated
by computing the determinant of a submatrix of the Laplacian
matrix corresponding to $F_{n}(x,y)$, which is obtained by removing a row and a
column corresponding to an arbitrary node of $F_{n}(x,y)$. The elements of the
Laplacian matrix associated with $F_{n}(x,y)$, represented by $F_n$, are
defined as follows: the non-diagonal entry $f_{ij}$ $(i\neq j)$ is
$-1$ (or $0$) if nodes $i$ and $j$ are (or not) directly connected by an edge,
while the diagonal entry $f_{ii}$ equals the degree of node $i$.

To find $N_{\rm ST}(n)$, we denote by $G_n$ the
submatrix of $F_n$ obtained by removing from $F_n$ the row and column corresponding to an initial node of
$F_{n}(x,y)$. Then, we have
\begin{equation}\label{f1}
N_{\rm ST}(n) = \det(G_n).
\end{equation}
Our aim is to determine $\det(G_n)$. Although for a general network of large size, the evaluation of its determinant is very hard and even impossible, below we show that for $F_{n}(x,y)$ and other self-similar networks, we can evaluate the associated determinants and derive closed-form expressions for the number of their spanning trees.

First, we define another submatrix of $F_n$, which is obtained by deleting the two rows and columns from
$F_n$ corresponding to the two initial nodes, and denote it by $H_n$. In what follows, we show the recursive relations for $\det(G_n)$ and $\det(H_n)$, thereby, expressing $\det(G_{n+1})$ and $\det(H_{n+1})$ in terms of $\det(G_n)$ and $\det(H_n)$, based on which we offer explicit solutions for these two quantities. For simplicity, let $f_n$, $g_n$, and $h_n$ represent $\det(F_n)$,
$\det(G_n)$, and $\det(H_n)$, respectively. Obviously, $f_n=\det(F_n)=0$ for all $n$, since the sum of each row in $F_n$ is zero.

It should be noted that the technique we develop for evaluating the determinants can be adapted to a wide class of self-similar networks other than the $(x,y)$-flowers. Here, we illustrate the computation process with the $(x,y)$-flowers. Moreover, we only consider in detail two particular and representative networks in the $(x,y)$-flower family, namely, $(1,3)$-flower and $(2,2)$-flower. For the general $(x,y)$-flowers we can treat them analogously although the computation is somewhat complex. In so doing, for notational simplicity,
we only provide the recursion relations governing $g_n$ and $h_n$ for the general $(x,y)$-flowers, but omit their detail computations.

\subsection{The $(1,3)$-flower}

According to the second construction algorithm of the networks, see Fig.~\ref{Alternative}, the Laplacian matrix of $F_{n+1}(1,3)$, denoted by $F_{n+1}$ in the case without confusion, can be expressed in terms of $H_{n}$ as
\begin{eqnarray}\label{Lap13}
F_{n+1} = \left(\begin{array}{cccccccc} 2d_n & -1 & 0 & -1 & p_n &
\emph{0} & \emph{0} & q_n \\ -1 & 2d_n & -1 & 0 & q_n & p_n & \emph{0} &
\emph{0} \\ 0 & -1 & 2d_n & -1 & \emph{0} & q_n & p_n & \emph{0}
\\ -1 & 0 & -1 & 2d_n & \emph{0} & \emph{0} & q_n & p_n \\
p^{\top}_n & q^{\top}_n & \emph{0}^{\top} & \emph{0}^{\top} & H_n & O & O & O \\
\emph{0}^{\top} & p^{\top}_n & q^{\top}_n & \emph{0}^{\top} & O & H_n & O & O
\\ \emph{0}^{\top} & \emph{0}^{\top} & p^{\top}_n & q^{\top}_n & O & O & H_n &
O \\ q^{\top}_n & \emph{0}^{\top} & \emph{0}^{\top} & p^{\top}_n & O & O
& O & H_n
\end{array}\right).
\end{eqnarray}
In Eq.~(\ref{Lap13}), the first two rows and two columns correspond to the two initial nodes of $F_{n+1}(1,3)$; 
$d_n$ is the degree of either of initial nodes at generation $n$; $p_n$ and $q_n$ are two row vectors with order $N_n-2=\frac{2}{3}(4^n-1)$ describing respectively the interaction between the two initial nodes 
and the other nodes in different replicas of $F_{n}(1,3)$ that are joined together forming $F_{n+1}(1,3)$; and the superscript $\top$ of a vector represents its transpose. The third and forth rows and columns describe the connection relation of the two nodes created at $n=1$ (each of which has the same degree $2d_n$ as the two initial nodes) and other nodes in $F_{n+1}(1,3)$.

Based on Eq.~(\ref{Lap13}), we can also represent $G_{n+1}$ and $H_{n+1}$ in terms of $H_{n}$, as
\begin{eqnarray}\label{f3}
G_{n+1}=\left(\begin{array}{ccccccc} 2d_n & -1 & 0 & q_n & p_n & 0 & 0
\\ -1 & 2d_n & -1 & 0 & q_n & p_n & 0 \\ 0 & -1 & 2d_n & 0 & 0 & q_n & p_n
\\ q^{\top}_n & 0^{\top} & 0^{\top} & H_n & O & O & O \\ p^{\top}_n &
q^{\top}_n & 0^{\top} & O & H_n & O & O \\ 0^{\top} & p^{\top}_n &
q^{\top}_n & O & O & H_n & O \\ 0^{\top} & 0^{\top} & p^{\top}_n & O & O
& O & H_n \end{array}\right)
\end{eqnarray}
and
\begin{eqnarray}\label{f4}
H_{n+1}=\left(\begin{array}{cccccc} 2d_n & -1 & 0 & q_n & p_n & 0
\\ -1 & 2d_n & 0 & 0 & q_n & p_n
\\ 0^{\top} & 0^{\top} & H_n & O & O & O \\ q^{\top}_n & 0^{\top} & O & H_n & O & O
\\ p^{\top}_n & q^{\top}_n & O & O & H_n & O \\ 0^{\top} & p^{\top}_n & O & O
& O & H_n \end{array}\right).
\end{eqnarray}

In the case of $n=2$, for example, Eqs.~(\ref{f3}) and (\ref{f4}) become respectively
\begin{eqnarray*}\label{f5}
G_2=\left(\begin{array}{ccccccc} 4 & -1 & 0 & q_1 & p_1 & 0 & 0
\\ -1 & 4 & -1 & 0 & q_1 & p_1 & 0 \\ 0 & -1 & 4 & 0 & 0 & q_1 & p_1
\\ q^{\top}_1 & 0^{\top} & 0^{\top} & H_1 & O & O & O \\ p^{\top}_1 &
q^{\top}_1 & 0^{\top} & O & H_1 & O & O \\ 0^{\top} & p^{\top}_1 &
q^{\top}_1 & O & O & H_1 & O \\ 0^{\top} & 0^{\top} & p^{\top}_1 & O & O
& O & H_1 \end{array}\right)
\end{eqnarray*}
and
\begin{eqnarray*}\label{f6}
H_2=\left(\begin{array}{cccccc} 4 & -1 & 0 & q_1 & p_1 & 0
\\ -1 & 4 & 0 & 0 & q_1 & p_1
\\ 0^{\top} & 0^{\top} & H_1 & O & O & O \\ q^{\top}_1 & 0^{\top}_1 & O & H_1 & O & O
\\ p^{\top}_1 & q^{\top}_1 & O & O & H_1 & O \\ 0^{\top} & p^{\top}_1 & O & O
& O & H_1 \end{array}\right),
\end{eqnarray*}
where $p_1 = \left(\begin{array}{cc}0 & -1\end{array}\right)$, $q_1 = \left(\begin{array}{cc}-1 & 0\end{array}\right)$, and $H_1 = \left(\begin{array}{cc} 2 & -1 \\ -1 & 2 \end{array}\right)$.

With the expressions for $G_{n+1}$ and $H_{n+1}$ given in Eqs.~(\ref{f3}) and~(\ref{f4}), we can evaluate the determinants of $G_n$ and $H_n$, namely, $g_n$ and $h_n$, on the basis of their recursive relations that will be derived.
We first show a detailed computation process for $g_n$, while the evaluation for $h_n$ can be dealt with in an analogous way.

Note that the first row of $G_{n+1}$ can be viewed as the sum, i.e., a linear combination with scalars being 1, of two vectors; that is,
\begin{eqnarray*}\label{f10}
&\quad&\left(\begin{array}{ccccccc} 2d_n & -1 & 0 & q_n & p_n & 0 &
0\end{array}\right)\nonumber\\
&=& \left(\begin{array}{ccccccc} d_n & 0 & 0 & q_n & 0 & 0 &
0\end{array}\right) + \left(\begin{array}{ccccccc} d_n & -1 & 0 & 0
& p_n & 0 &0\end{array}\right).
\end{eqnarray*}
Analogously, the second and third rows can also be regarded separately as the sum of two vectors, namely,
\begin{eqnarray*}\label{f11}
&\quad&\left(\begin{array}{ccccccc} -1 & 2d_n & -1 & 0 & q_n & p_n &
0\end{array}\right)\\
&=& \left(\begin{array}{ccccccc} -1 & d_n & 0 & 0 & q_n & 0 &
0\end{array}\right) + \left(\begin{array}{ccccccc} 0 & d_n & -1 & 0
& 0 & p_n & 0\end{array}\right),
\end{eqnarray*}
and
\begin{eqnarray*}\label{f12}
&\quad&\left(\begin{array}{ccccccc} 0 & -1 & 2d_n & 0 & 0 & q_n &
p_n\end{array}\right)\\
&=& \left(\begin{array}{ccccccc} 0 & -1 & d_n & 0 & 0 & q_n &
0\end{array}\right) + \left(\begin{array}{ccccccc} 0 & 0 & d_n & 0 &
0 & 0 & p_n\end{array}\right).
\end{eqnarray*}
Thus, according to the properties of determinants, $g_{n+1}$ can be decomposed as
the sum of eight determinants, as follows:
\begin{widetext}
\begin{eqnarray*}\label{f13}
g_{n+1}&=&\left|\begin{array}{ccccccc} d_n & 0 & 0 & q_n & 0 & 0 & 0
\\ -1 & d_n & 0 & 0 & q_n & 0 & 0 \\ 0 & -1 & d_n & 0 & 0 & q_n & 0
\\ q^{\top}_n & 0^{\top} & 0^{\top} & H_n & O & O & O \\ p^{\top}_n &
q^{\top}_n & 0^{\top} & O & H_n & O & O \\ 0^{\top} & p^{\top}_n &
q^{\top}_n & O & O & H_n & O \\ 0^{\top} & 0^{\top} & p^{\top}_n & O & O
& O & H_n \end{array}\right| + \left|\begin{array}{ccccccc} d_n & 0
& 0 & q_n & 0 & 0 & 0
\\ -1 & d_n & 0 & 0 & q_n & 0 & 0 \\ 0 & 0 & d_n & 0 & 0 & 0 & p_n
\\ q^{\top}_n & 0^{\top} & 0^{\top} & H_n & O & O & O \\ p^{\top}_n &
q^{\top}_n & 0^{\top} & O & H_n & O & O \\ 0^{\top} & p^{\top}_n &
q^{\top}_n & O & O & H_n & O \\ 0^{\top} & 0^{\top} & p^{\top}_n & O & O
& O & H_n \end{array}\right| + \left|\begin{array}{ccccccc} d_n & 0
& 0 & q_n & 0 & 0 & 0
\\ 0 & d_n & -1 & 0 & 0 & p_n & 0 \\ 0 & -1 & d_n & 0 & 0 & q_n & 0
\\ q^{\top}_n & 0^{\top} & 0^{\top} & H_n & O & O & O \\ p^{\top}_n &
q^{\top}_n & 0^{\top} & O & H_n & O & O \\ 0^{\top} & p^{\top}_n &
q^{\top}_n & O & O & H_n & O \\ 0^{\top} & 0^{\top} & p^{\top}_n & O & O
& O & H_n \end{array}\right|\nonumber\\
&\quad&+\left|\begin{array}{ccccccc} d_n & 0 & 0 & q_n & 0 & 0 & 0
\\ 0 & d_n & -1 & 0 & 0 & p_n & 0 \\ 0 & 0 & d_n & 0 & 0 & 0 & p_n
\\ q^{\top}_n & 0^{\top} & 0^{\top} & H_n & O & O & O \\ p^{\top}_n &
q^{\top}_n & 0^{\top} & O & H_n & O & O \\ 0^{\top} & p^{\top}_n &
q^{\top}_n & O & O & H_n & O \\ 0^{\top} & 0^{\top} & p^{\top}_n & O & O
& O & H_n \end{array}\right| + \left|\begin{array}{ccccccc} d_n & -1
& 0 & 0 & p_n & 0 & 0
\\ -1 & d_n & 0 & 0 & q_n & 0 & 0 \\ 0 & -1 & d_n & 0 & 0 & q_n & 0
\\ q^{\top}_n & 0^{\top} & 0^{\top} & H_n & O & O & O \\ p^{\top}_n &
q^{\top}_n & 0^{\top} & O & H_n & O & O \\ 0^{\top} & p^{\top}_n &
q^{\top}_n & O & O & H_n & O \\ 0^{\top} & 0^{\top} & p^{\top}_n & O & O
& O & H_n \end{array}\right|+\left|\begin{array}{ccccccc} d_n & -1 &
0 & 0 & p_n & 0 & 0
\\ -1 & d_n & 0 & 0 & q_n & 0 & 0 \\ 0 & 0 & d_n & 0 & 0 & 0 & p_n
\\ q^{\top}_n & 0^{\top} & 0^{\top} & H_n & O & O & O \\ p^{\top}_n &
q^{\top}_n & 0^{\top} & O & H_n & O & O \\ 0^{\top} & p^{\top}_n &
q^{\top}_n & O & O & H_n & O \\ 0^{\top} & 0^{\top} & p^{\top}_n & O & O
& O & H_n \end{array}\right|\nonumber\\
&\quad&+\left|\begin{array}{ccccccc} d_n & -1 & 0 & 0 & p_n & 0 & 0
\\ 0 & d_n & -1 & 0 & 0 & p_n & 0 \\ 0 & -1 & d_n & 0 & 0 & q_n & 0
\\ q^{\top}_n & 0^{\top} & 0^{\top} & H_n & O & O & O \\ p^{\top}_n &
q^{\top}_n & 0^{\top} & O & H_n & O & O \\ 0^{\top} & p^{\top}_n &
q^{\top}_n & O & O & H_n & O \\ 0^{\top} & 0^{\top} & p^{\top}_n & O & O
& O & H_n \end{array}\right| + \left|\begin{array}{ccccccc} d_n & -1
& 0 & 0 & p_n & 0 & 0
\\ 0 & d_n & -1 & 0 & 0 & p_n & 0 \\ 0 & 0 & d_n & 0 & 0 & 0 & p_n
\\ q^{\top}_n & 0^{\top} & 0^{\top} & H_n & O & O & O \\ p^{\top}_n &
q^{\top}_n & 0^{\top} & O & H_n & O & O \\ 0^{\top} & p^{\top}_n &
q^{\top}_n & O & O & H_n & O \\ 0^{\top} & 0^{\top} & p^{\top}_n & O & O
& O & H_n \end{array}\right|.
\end{eqnarray*}
\end{widetext}

Therefore, the evaluation of $g_{n+1}$ is reduced to finding the values of the eight determinants on the right-hand side of Eq.~(\ref{f13}), which are denoted sequentially as $g_{n+1}^{(i)}$ $(1 \le i \le 8)$. Using some elementary matrix operations, we can verify that $g_{n+1}^{(1)} = g_{n+1}^{(2)}
= g_{n+1}^{(4)} = g_{n+1}^{(8)}$ and $g_{n+1}^{(3)} = g_{n+1}^{(5)}
= g_{n+1}^{(6)} = g_{n+1}^{(7)}$. 
Having $g_{n+1}$ in terms of the quantities of $g_{n+1}^{(i)}$, the next step is to explicitly determine these quantities. In view of the equivalence of some $g_{n+1}^{(i)}$, we only compute $g_{n+1}^{(2)}$ and $g_{n+1}^{(3)}$.

For $g_{n+1}^{(2)}$ and $g_{n+1}^{(3)}$, utilizing some elementary matrix operations, we can easily obtain
\begin{eqnarray*}\label{f14}
g_{n+1}^{(2)}=h_n\left|\begin{array}{cccccc}d_n & q_n & 0 & \emph{0} &
0 & \emph{0} \\ q^{\top}_n & H_n & \emph{0}^{\top} & O &
\emph{0}^{\top} & O \\ -1 & \emph{0} & d_n & q_n & 0 & \emph{0} \\
p^{\top}_n & O & q^{\top}_n & H_n & \emph{0}^{\top} & O \\ 0 & \emph{0}
& 0 & \emph{0} & d_n & p_n \\ \emph{0}^{\top} & O & \emph{0}^{\top} &
O & p^{\top}_n & H_n \end{array}\right| = h_n(g_n)^{3}
\end{eqnarray*}
and
\begin{eqnarray*}\label{f15}
g_{n+1}^{(3)} &=& (h_n)^2\left|\begin{array}{ccccc}d_n & -1 & p_n & 0 & \emph{0} \\
-1 & d_n & q_n & 0 & \emph{0} \\ p^{\top}_n & q^{\top}_n & H_n &
\emph{0}^{\top} & O \\ 0 & 0 & \emph{0} & d_n & q_n \\ \emph{0}^{\top}
& \emph{0}^{\top} & O & q^{\top}_n & H_n \end{array}\right| \nonumber \\
&=&(h_n)^2f_ng_n=0.
\end{eqnarray*}
Plugging these expressions into Eq.~(\ref{f13}), we have
\begin{eqnarray}\label{f16}
g_{n+1}=\sum_{i=1}^{8}g_{n+1}^{(i)} = 4h_n(g_n)^3.
\end{eqnarray}

Next, we derive the recursion relation for $h_{n+1}$. Through a computational process similar to that for $g_{n+1}$, we obtain the following relation:
\begin{widetext}
\begin{eqnarray*}\label{f18}
h_{n+1}&=&h_n\left(\left|\begin{array}{ccccc}d_n & 0 & q_n & \emph{0}
& \emph{0} \\ -1 & d_n & \emph{0} & q_n & \emph{0} \\ q^{\top}_n &
\emph{0}^{\top} & H_n & O & O \\ p^{\top}_n & l^{\top}_n & O & H_n & O
\\ \emph{0}^{\top} & p^{\top}_n & O & O & H_n
\end{array}\right| + \left|\begin{array}{ccccc}d_n & 0 & q_n & \emph{0}
& \emph{0} \\ 0 & d_n & \emph{0} & \emph{0} & p_n \\ q^{\top}_n &
\emph{0}^{\top} & H_n & O & O \\ p^{\top}_n & q^{\top}_n & O & H_n & O
\\ \emph{0}^{\top} & p^{\top}_n & O & O & H_n
\end{array}\right| + \left|\begin{array}{ccccc}d_n & -1 & \emph{0} &
p_n & \emph{0} \\ -1 & d_n & \emph{0} & q_n & \emph{0} \\ q^{\top}_n &
\emph{0}^{\top} & H_n & O & O \\ p^{\top}_n & q^{\top}_n & O & H_n & O
\\ \emph{0}^{\top} & p^{\top}_n & O & O & H_n
\end{array}\right| + \left|\begin{array}{ccccc}d_n & -1 & \emph{0} &
p_n & \emph{0} \\ 0 & d_n & \emph{0} & \emph{0} & p_n \\ q^{\top}_n &
\emph{0}^{\top} & H_n & O & O \\ p^{\top}_n & q^{\top}_n & O & H_n & O
\\ \emph{0}^{\top} & p^{\top}_n & O & O & H_n
\end{array}\right|\right)\nonumber\\
&=& 3(h_n)^2(g_n)^2+(h_n)^3f_n = 3(h_n)^2(g_n)^2.
\end{eqnarray*}
\end{widetext}

Having obtained the above recursive expressions for $g_{n+1}$ and $h_{n+1}$, we are now in a position to explicitly determine $g_{n}$. For this purpose, we examine a new quantity $g_n/h_n$, which obviously satisfies the following recursive relation:
\begin{equation*}\label{f19}
\frac{g_{n+1}}{h_{n+1}} = \frac{4}{3}\frac{g_n}{h_n}.
\end{equation*}
Considering the initial condition $g_1/h_1 = 4/3$, the above equation is
solved by induction to yield
\begin{equation*}\label{f20}
\frac{g_n}{h_n} = \left(\frac{4}{3}\right)^{n}.
\end{equation*}
Hence,
\begin{equation}\label{f21}
h_n = \left(\frac{3}{4}\right)^n g_n.
\end{equation}
Substituting Eq.~(\ref{f21}) into Eq.~(\ref{f16}) and using $g_1 =
4$, we can solve Eq.~(\ref{f16}) to obtain the exact
number of spanning trees:
\begin{equation}\label{f22}
N_{\rm ST}(n) = g_n = 3^{(4^n-3n-1)/9}4^{(2\times 4^{n}+3n-2)/9}.
\end{equation}

It is easy to express $N_{\rm ST}(n)$ in terms of the
network size $N_n$, so as to obtain the relation governing the
two quantities. Recalling that for $(1,3)-$flower, $N_n = (2\times
4^n+4)/3$, hence we have
$4^{n}=(3N_n-4)/2$ and $n=[\ln(3N_n-4)/\ln2-1]/2$. These relations allow us to
recast $N_{\rm ST}(n)$ as a function of $N_n$:
\begin{equation*}\label{ST04}
N_{\rm ST}(n)=3^{[N_n-\ln(3N_n-4)/\ln2-1]/6}4^{[2N_n+\ln(3N_n-4)/\ln2-5]/6}\,,
\end{equation*}
which indicates that $N_{\rm ST}(n)$ approximately increases exponentially in $N_n$.

Then we can define the entropy of spanning trees for the $(1,3)-$flower as
the limiting value~\cite{BuPe93,Ly05,LyPeSc08}:
\begin{equation*}\label{e1}
E_{\rm{(1,3)-flower}} =
\lim_{N_n\rightarrow\infty}\frac{\ln{N_{\rm ST(n)}}}{N_n}=\frac{1}{6}(4\ln2+\ln3)\approx0.6452.
\end{equation*}

\subsection{The $(2,2)$-flower}

We proceed to determine the number of spanning trees in the
$(2,2)$-flower. In the computation process, we use the same notations as those defined for the $(1,3)$-flower. Similarly to the $(1,3)$-flower, we can derive the recursion relations for $G_{n+1}$ and $H_{n+1}$, as
\begin{eqnarray}\label{f23}
G_{n+1}=\left(\begin{array}{ccccccc} 2d_n & 0 & 0 & s_n & r_n & 0 & 0
\\ 0 & 2d_n & 0 & 0 & s_n & r_n & 0 \\ 0 & 0 & 2d_n & 0 & 0 & s_n & r_n
\\ s^{\top}_n & 0^{\top} & 0^{\top} & H_n & O & O & O \\ r^{\top}_n &
s^{\top}_n & 0^{\top} & O & H_n & O & O \\ 0^{\top} & r^{\top}_n &
s^{\top}_n & O & O & H_n & O \\ 0^{\top} & 0^{\top} & r^{\top}_n & O & O
& O & H_n \end{array}\right),
\end{eqnarray}
and
\begin{eqnarray}\label{f24}
H_{n+1}=\left(\begin{array}{cccccc} 2d_n & 0 & s_n & r_n & 0 & 0
\\ 0 & 2d_n & 0 & 0 & s_n & r_n
\\ s^{\top}_n & 0^{\top} & H_n & O & O & O \\ r^{\top}_n & 0^{\top} & O & H_n & O & O
\\ 0^{\top} & s^{\top}_n & O
& O & H_n & O \\ 0^{\top} & r^{\top}_n & O & O & O & H_n
\end{array}\right),
\end{eqnarray}
respectively. In Eqs.~(\ref{f23}) and~(\ref{f24}), $r_n$ and $s_n$ are two vectors describing the connection relation between an initial node and the other nodes.

Based on the above two relations, we obtain the following relations for the determinants corresponding to $G_{n+1}$ and $H_{n+1}$ by using some elementary matrix operations:
\begin{equation*}\label{f25}
g_{n+1} = 4h_n(g_n)^3 + 4(h_n)^2f_n g_n = 4h_n(g_n)^3,
\end{equation*}
and
\begin{equation*}\label{f26}
h_{n+1} = 4(h_n)^2(g_n)^2.
\end{equation*}
These two recursion relations, coupled with the initial conditions $g_1 = 4$ and $h_1 = 4$, yield
\begin{equation}\label{f27}
N_{\rm ST}(n) = g_n = 2^{\frac{2}{3}(4^n-1)},
\end{equation}
which can be further expressed in terms of the network size $N_n$ as
\begin{eqnarray*}\label{f28}
N_{\rm ST}(n)= 2^{N_n-2}\,.
\end{eqnarray*}
Thus, for the $(2,2)$-flower, the number of spanning trees also scales exponentially with the network size. And its entropy of spanning trees is
\begin{equation*}\label{e2}
E_{\rm{(2,2)-flower}}=\lim_{N_n\rightarrow\infty}\frac{\ln{N_{ST}(n)}}{N_n}
=\ln2\approx0.6931\,,
\end{equation*}
a value larger than that corresponding to the $(1,3)$-flower.

\subsection{The $(x,y)$-flowers}

The above process for computing spanning trees in the $(1,3)$-flower and the $(2,2)$-flower can be used to determine the number of spanning trees in the whole family of $(x,y)$-flowers. Since for the general $(x,y)$-flowers, the recursive relations for $G_{n+1}$ and $H_{n+1}$ are lengthy and the computation process for $g_{n+1}$ and $h_{n+1}$ is more complex, we only provide the recursive relations for $g_{n+1}$ and $h_{n+1}$, leaving out the derivation detail, which is similar to those for the two particular cases considered above. The recursive relations of $g_{n+1}$ and $h_{n+1}$ for the $(x,y)$-flowers are given by
\begin{equation}\label{f31}
g_{n+1}=(x+y)h_n(g_n)^{x+y-1}
\end{equation}
and
\begin{equation}\label{f32}
h_{n+1}=x y(h_n)^2(g_n)^{x+y-2}.
\end{equation}

Considering the initial conditions $g_1 = x+y$ and $h_1 = x y$, we can solve
Eqs.~(\ref{f31}) and~(\ref{f32}) to obtain the exact number of spanning trees for
the $(x,y)$-flowers:
\begin{eqnarray}\label{f33}
N_{\rm ST}(n) &=& g_n = (x y)^{\frac{(x+y)^n - (x+y-1)n -
1}{(x+y-1)^2}}\nonumber\\
&\quad&\times(x+y)^{\frac{(x+y-2)(x+y)^n + (x+y-1)n
-(x+y-2)}{(x+y-1)^2}}.
\end{eqnarray}
For $x=1$ and $y=2$, Eq.~(\ref{f33}) recovers the result previously obtained in~\cite{ZhLiWuZh10}; for $x=1$ and $y=3$, Eq.~(\ref{f33}) reduces to Eq.~(\ref{f22}); while for $x=2$ and $y=2$, Eq.~(\ref{f33}) agrees with the result given by Eq.~(\ref{f27}). The consistency confirms the validity of the formula for spanning trees in the $(x,y)$-flowers.

Since the size for the $(x,y)$-flowers is $N_n = (x+y-2)/(x+y-1)(x+y)^n+(x+y)/(x+y-1)$, it is evident from Eq.~(\ref{f33}) that $N_{\rm ST}(n)$ scales exponentially with $N_n$. This permits to get the entropies of spanning trees
for the $(x,y)$-flowers as follows:
\begin{eqnarray}\label{e3}
E_{(x,y)-{\rm flowers}}&=&\lim_{N_n\rightarrow\infty}\frac{\ln{N_{\rm ST}(n)}}{N_n}\nonumber\\
&=& \frac{\ln{x y}+(x+y-2)\ln{(x+y)}}{(x+y-1)(x+y-2)}.
\end{eqnarray}

It is worthwhile to mention that the above technique and process for finding spanning trees in the $(x,y)$-flowers by evaluating determinants can be adapted to other self-similar networks. In order to show the universality, in the Appendix we apply the approach to determine the number of spanning trees and their entropies for a small-world network~\cite{BaCoDaFi09} with an exponential degree distribution and a scale-free Apollonian network~\cite{AnHeAnSi05,ZhChZhFaGuZo08}, and obtain explicit solutions to the corresponding quantities.

\begin{figure}
\begin{center}
\includegraphics[width=0.9\linewidth,trim=0 0 0 0]{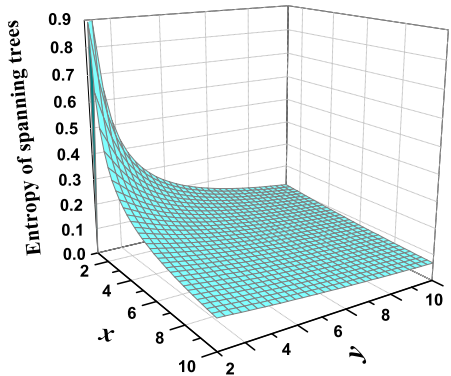}
\caption{(Color online) The entropies of spanning trees in the $(x,y)$-flowers, depending on the two parameters $x$ and $y$.} \label{Entro}
\end{center}
\end{figure}

\subsection{Analysis of the results}

The entropy for spanning trees is a very important quantity
characterizing the network structure. Thus, it is interesting to
show its dependence on the network structure. Since, for the whole family of the $(x,y)$-flowers, its architecture is controlled by the two parameters $x$ and $y$, it is expected that the entropy is dependent on $x$ and $y$. In Fig.~\ref{Entro}, we show how the entropy varies with $x$ and $y$. The intrinsic dependence relation of entropy on $x$ and $y$ is somewhat complex. Here we are mainly concerned about how the entropy  $E_{(x,y)-{\rm flowers}}$ is reliant on $x$ and $y$ when the sum $x+y$ is fixed with $x+y\geq 4$. The main reason why we are interested in this case lies in that for this case it is meaningful to compare the entropy. As shown in section~\ref{SecModel}, for the $(x,y)$-flowers with fixed $x+y$, they have an identical average degree. In addition, for fixed $x+y$, the networks have the same degree sequence and thus the same degree distribution. Particularly, for $x > 1$ with fixed $x+y$, the networks are fractal having the fractal dimension monotonously decreasing in $x$.

Figure~\ref{Entro} shows that when $x+y$ is fixed, $E_{(x,y)-{\rm flowers}}$ is determined by the difference between $x$ and $y$: the larger the difference, the smaller the entropy. Thus, in the case of fixed $x+y$, $E_{(x,y)-{\rm flowers}}$ is an increasing function of parameter $x$. For example, as shown above, when $x+y=4$, the entropy for the $(1,3)$-flower is less than that for the $(2,2)$-flower. Thus, we can conclude that the power-law degree distribution alone can not determine the entropies for spanning trees in scale-free networks. In fact, other than scale-free networks, the degree distributions of other networks do not suffice to characterize the number of spanning trees either. For instance, both the small-world network~\cite{BaCoDaFi09} and the Koch network family~\cite{ZhZhXiChLiGu09,ZhGaChZhZhGu10}, to be addressed in the Appendix, have the same average degree and the same entropy of spanning trees, but their degree distributions are disparate: the former is exponential while the latter is scale-free.

As pointed out above, degree distribution alone cannot determine the number of spanning trees. Then an important question arises: which structural property has the decisive effect on the number of spanning tress in the $(x,y)$-flower family with an  identical value of $x+y$? Since, for the whole family of $(x,y)$-flowers with the same $x+y$, they have the same average node degree, the same degree sequence and thus the same degree distribution, and the same (zero) clustering coefficient, we argue that the difference in the number of spanning trees seen in Fig.~\ref{Entro} is attributed solely to the fractality. Specifically, the fractal property of networks significantly increases the number of spanning trees, which is related to the fractal dimension: the larger the fractal dimension, the less the number of spanning trees. This is obvious from Fig.~\ref{Entro}. For those $(x,y)$-flowers with the same value of $x+y$, when $x$ grows from 1 to its maximum, their fractal dimension $d_{\rm f}$ decreases from $\infty$ to its minimum, while the entropies of spanning trees increase from their minimum to maximum. Particularly, when $x=1$, the network is non-fractal, it has the minimum number of spanning trees.

Before closing this section, we discuss the case of different values of $x+y$. Figure~\ref{Entro} shows that for two network classes with different values of $x+y$, the class with smaller $x+y$ has a larger entropy than that of the other class. This can be easily understood. Since for two networks with distinct values of $x+y$, their average node degrees are also disparate: the network with smaller $x+y$ has a larger average node degree than that the other. Thus, the edge density of the former is greater than that of the latter, which is responsible for the difference of entropies for spanning trees in the two networks.

\section{conclusions}

In this paper, we have developed a rather generic technique for evaluating determinants used to compute the numbers of spanning trees in self-similar networks. Our method is grounded on the recursive relations for the determinants of submatrices of the Laplacian matrix corresponding to the network at two consecutive generations, which are established based on the structure of the network. An advantage of our approach is that it avoids the laborious computation of a determinant and reduces the computational complexity that a standard method needs. We have illustrated our technique by applying it to the family of $(x,y)$-flowers, which exhibits rich and special structural properties, as well as some remarkable features of many real systems. Moreover, to demonstrate the universality of our approach, we have applied it to some other networks with self-similarity. We obtained explicit formulas for the numbers of spanning trees in different networks. Our results show that degree distribution alone does not suffice to characterize the number of spanning trees in a network, and that fractal dimension has a predominant influence on the  number of spanning trees in fractal scale-free networks with the same average degree.

\begin{acknowledgments}
This work was supported by the National Natural Science Foundation
of China under Grant No. 61074119 and by the Hong Kong Research Grants
Council under the GRF Grant CityU 1117/10E.
\end{acknowledgments}

\appendix*

\section{Spanning trees in other self-similar networks}

To illustrate the university of our proposed method, we apply it to calculate the number of spanning trees in some other self-similar networks with different degree distributions, exponential or scale-free.

\subsection{Small-world exponential network}

\begin{figure}
\begin{center}
\includegraphics[width=0.7\linewidth,trim=0 0 0 0]{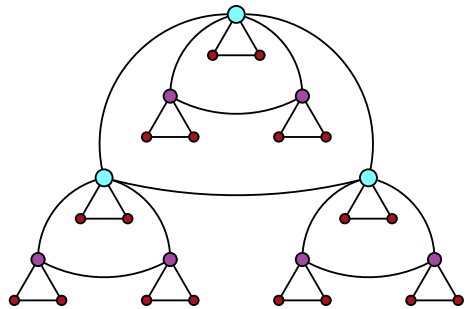}
\caption{(Color online) An illustration for the network $S_{2}$.} \label{SmallW}
\end{center}
\end{figure}

We first discuss a self-similar small-world network with an exponential form of degree distribution~\cite{BaCoDaFi09}, which is observed from some real-life systems~\cite{AmScBaSt00}. The small-world network is constructed iteratively~\cite{BaCoDaFi09}, denoted by $S_{n}$
after $n\quad(n\ge0)$ iterations. The network starts from a triangle $S_{0}$, with three nodes called initial nodes. For $n \ge 1$, $S_{n}$ is obtained from $S_{n-1}$: for each existing node in $S_{n-1}$, two new nodes are generated, which and their mother node together form a new triangle. Figure~\ref{SmallW} shows the structure of $S_{2}$. It is easy to derive that the total number of nodes in $S_{n}$ is $N_n=3^{n+1}$. The resultant network has a degree distribution decaying exponentially with the degree. In addition, it is small-world with the average distance growing logarithmically with its size.

We now begin to evaluate the number of spanning trees in $S_{n}$. In the case without any confusion, we use $F_{n}$ to denote the Laplacian matrix for $S_{n}$, and use  $G_{n}$ to denote a submatrix of $F_{n}$, which is obtained by removing from $F_{n}$ the row and column corresponding to one of three initial nodes of $S_{n}$. We have
\begin{eqnarray}\label{New2}
F_{n+1}=\left(\begin{array}{cccccc}
d_{n}+2 & -1 & -1 & u_n & \emph{0} & \emph{0} \\
-1 & d_{n}+2 & -1 & \emph{0} & u_n & \emph{0} \\
-1 & -1 & d_{n}+2 & \emph{0} & \emph{0} & u_n \\
u_n^{\top} & \emph{0}^{\top} & \emph{0}^{\top} & G_{n} & O & O \\
\emph{0}^{\top} & u_n^{\top}  &\emph{0}^{\top} & O & G_{n} & O \\
\emph{0}^{\top} & \emph{0}^{\top} & u_n^{\top} & O & O & G_{n}\end{array}\right)
\end{eqnarray}
and
\begin{eqnarray}\label{X2}
G_{n+1}=\left(\begin{array}{ccccc}
d_{n}+2 & -1 & \emph{0} & u_n & \emph{0} \\
-1 & d_{n}+2 & \emph{0} & \emph{0} & u_n \\
\emph{0}^{\top} & \emph{0}^{\top} & G_{n} & O & O \\
u_n^{\top} & \emph{0}^{\top} & O & G_{n} & O \\
\emph{0}^{\top} & u_n^{\top} & O &
O & G_{n}\end{array}\right)\,,
\end{eqnarray}
where $d_{n}$ is the degree of an initial node in $S_n$, and
$u_n$ is a row vector representing the interaction between an initial node
and the other non-initial nodes.

According to the Matrix Tree Theorem, the number of spanning trees in $S_n$
is $N_{ST}(n)={\rm det}(G_{n})$. Using a similar iterative process as that in the main text, Eq.~(\ref{X2}) leads to the following relation: ${\rm det}(G_{n})=3({\rm det}(G_{n-1}))^3$. Considering the initial condition of ${\rm det}(G_{0})=3$, we obtain the exact solution to the number of spanning trees in the small-world
network $S_n$ as
\begin{equation}\label{X11}
N_{\rm ST}(n)=3^{(3^{n+1}-1)/2}\,,
\end{equation}
and the entropy of spanning trees as
\begin{equation}\label{e4}
E_{S_n}
=\lim_{N_n\rightarrow\infty}\frac{\ln{N_{\rm ST}(n)}}{N_n}=\frac{1}{2}\ln{3}\approx0.5493.
\end{equation}

The above computational process for $S_n$ can also be applied to the scale-free Koch networks~\cite{ZhZhXiChLiGu09,ZhGaChZhZhGu10} with the same average node degree 3 as that of $S_n$. Interestingly, the spanning tree entropy for the Koch networks is identical to $S_n$ despite the fact that they have totally different types of degree distributions.

\subsection{Small-world scale-free Apollonian network}

\begin{figure}
\begin{center}
\includegraphics[width=.35\linewidth,trim=20 50 70 0]{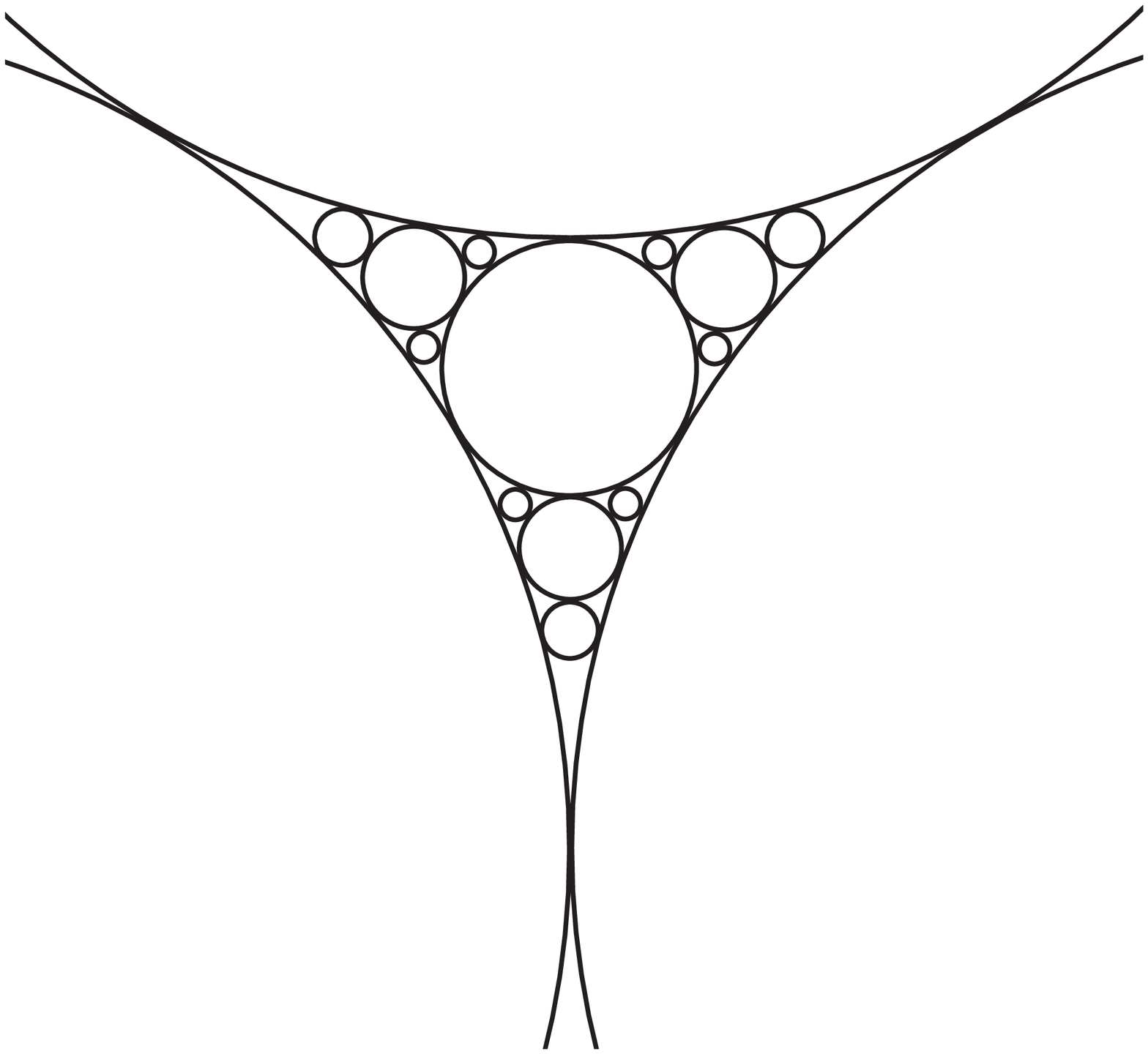}
\includegraphics[width=.48\linewidth,trim=70 50 70 0]{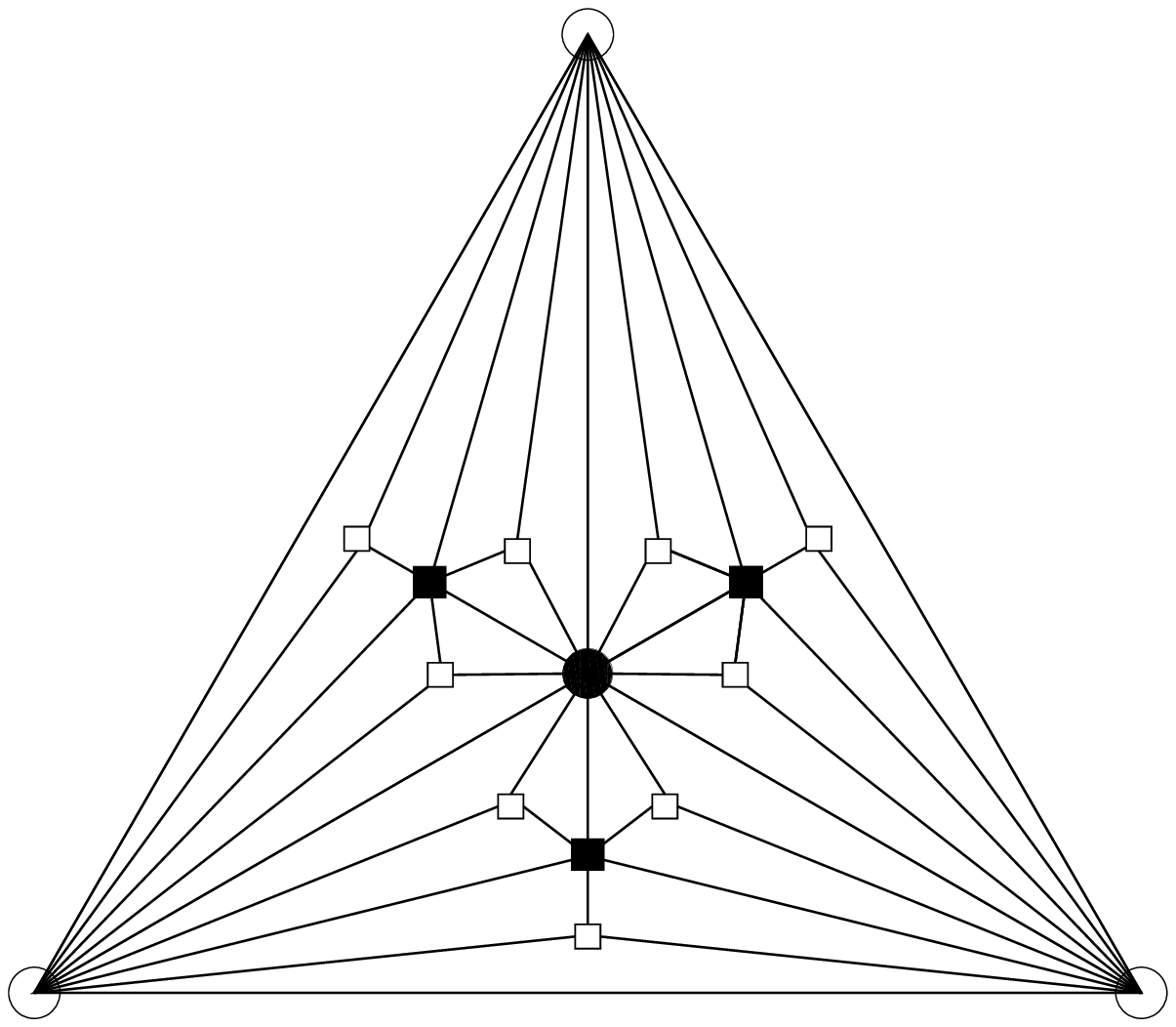}
\end{center}
\caption{ The classic Apollonian packing of disks and its
corresponding Apollonian network.} \label{Apollo}
\end{figure}

We apply our technique to determine the spanning trees in the Apollonian network with both small-world and scale-free properties~\cite{AnHeAnSi05,ZhChZhFaGuZo08}. The Apollonian network is
derived from the classic Apollonian packing. This well-known packing
problem~\cite{KaSu43} starts with three mutually touching disks and
their interstice is filled by a disk that touches all the three
initial disks, forming three smaller interstices to be filled, as
shown in the left panel of Fig.~\ref{Apollo}. Let each disk denote a
node located at the center of the corresponding disk, linking the
centers of touching disks by lines leads to the Apollonian network,
as displayed in the right panel of Fig.~\ref{Apollo}. The Apollonian
network can also be constructed iteratively and has the small-world
effects~\cite{ZhChZhFaGuZo08}. After $n$ iterations, the size of the
Apollonian network is $N_n=\frac{3^{n}+5}{2}$.

Making use of our technique for evaluating the corresponding determinant, we can determine the number of spanning trees in the Apollonian network. Since the computational process is complicated and the relation equations are lengthy, we only provide the main results.
The exact number of spanning trees in the Apollonian
network after $n$ iterations is
\begin{eqnarray}\label{ST05}
N_{\rm ST}(n)=\frac{1}{4}(3^n+5^n)^2
3^{(3^{n}-6n+3)/4}5^{(3^{n}-2n-1)/4}\,.
\end{eqnarray}
Consequently, the entropy for spanning trees in the Apollonian network is
\begin{equation}\label{Entropy03}
E_{\rm Apol}=\lim_{N_n \rightarrow \infty }\frac{\ln
N_{\rm ST}(n)}{N_n}=\frac{1}{2}(\ln3+\ln
5)\approx 1.3540\,,
\end{equation}
which is smaller than 1.9902, the entropy for spanning
trees in three-dimensional lattices~\cite{ChSh03} having the same
mean degree 6 as this Apollonian network.

\hfill
\nocite{*}


\begin{references}


\bibitem{BuPe93}
R. Burton and R. Pemantle, Ann. Probab. {\bf 21}, 1329 (1993).

\bibitem{BrMa97}
A. Z. Broder and E. W. Mayr,
J. Algorithms {\bf 24}, 171 (1997).

\bibitem{Ly05}
R. Lyons, Combin. Probab. Comput. {\bf 14}, 491 (2005).

\bibitem{LyPeSc08}
R. Lyons, R. Peled and O. Schramm,  Combin. Probab. Comput. {\bf
17}, 711 (2008).

\bibitem{Wu77}
F.-Y. Wu, J. Phys. A {\bf 10}, L113 (1977).

\bibitem{ShWu00}
R. Shrock and F.-Y. Wu,
J. Phys. A {\bf 33}, 3881 (2000).

\bibitem{ChSh03}
J. L. Felker and R. Lyons,
J. Phys. A {\bf 36}, 8361 (2003).

\bibitem{ChChYa07}
S.-C. Chang, L.-C. Chen, and W.-S. Yang, J. Stat. Phys. {\bf 126},
649 (2007).

\bibitem{TeWa10}
E. Teufl and S. Wagner, J. Phys. A {\bf 43}, 415001 (2010).

\bibitem{TeWa11}
E. Teufl and S. Wagner, J. Stat. Phys. {\bf 142},
879 (2011).


\bibitem{JaTh89}
R. Jayakumar and K. Thulasiraman. IEEE Trans. Circuits Syst. {\bf
36}, 219 (1989).


\bibitem{Bo86}
F. T. Boesch
J. Graph Theory {\bf 10}, 339 (1986). 

\bibitem{SzAlKe03}
G. J. Szab\'o, M. Alava, and J. Kert\'esz, Physica A {\bf 330}, 31
(2003).

\bibitem{BoSaSu09}
F. T. Boesch, A. Satyanarayana, and C. L. Suffel,
Networks {\bf 54}, 99 (2009).

\bibitem{Dh90}
D. Dhar, Phys. Rev. Lett. {\bf 64}, 1613 (1990).

\bibitem{Dh06}
D. Dhar
Physica A {\bf 369}, 29 (2006).

\bibitem{Me05}
C. Merino
Discrete Math. {\bf 302}, 188 (2005).

\bibitem{BaTaWi87}
P. Bak, C. Tang, and K. Wiesenfeld, Phys. Rev. Lett. {\bf 59}, 381
(1987).

\bibitem{GoSaKaKi06}
K.-I. Goh, G. Salvi, B. Kahng and D Kim,
Phys. Rev. Lett. {\bf 96}, 018701 (2006).

\bibitem{WuBrHaSt06}
Z. H. Wu, L. A. Braunstein, S. Havlin, and H. E. Stanley, Phys. Rev.
Lett. {\bf 96}, 148702 (2006).

\bibitem{BaGu10}
R. B. Bapat and S. Gupta,
Indian J. Pure Appl. Math. {\bf 41}, 1 (2010).

\bibitem{NoRi04}
J. D. Noh and H. Rieger, Phys. Rev. Lett. {\bf 92}, 118701 (2004).

\bibitem{CoBeTeVoKl07}
S. Condamin, O. B\'enichou, V. Tejedor, R. Voituriez, and J.
Klafter, Nature (London) {\bf 450}, 77 (2007).

\bibitem{DhDh97}
D. Dhar and A. Dhar
Phys. Rev. E {\bf 55}, 2093(R) (1997).

\bibitem{WuCh04}
B. Y. Wu and K.-M. Chao, \emph{Spanning Trees and Optimization
Problems} (Chapman $\&$ Hall/CRC, Boca Raton, 2004).

\bibitem{DoDu01}
R. Dobrin and P. M. Duxbury, Phys. Rev. Lett. {\bf 86}, 5076 (2001).

\bibitem{Wu02}
F.-Y. Wu, Int. J. Mod. Phys. B {\bf 16}, 1951 (2002).

\bibitem{KiNoJe04}
D.-H. Kim, J. D. Noh, and H. Jeong, Phys. Rev. E {\bf 70}, 046126
(2004).

\bibitem{MaAlBa05}
P. J. Macdonald, E. Almaas and A.-L. Barab\'asi, Europhys. Lett.
{\bf 72}, 308 (2005).

\bibitem{SeBoVe09}
M. A. Serrano, M. Boguna, and A. Vespignani,
Proc. Natl. Acad. Sci. U.S.A. {\bf 106}, 6483 (2009).

\bibitem{Kr10}
M. Krivelevich,
SIAM J. Discrete Math. {\bf 24}, 1495 (2010).

\bibitem{OzYa11}
K. Ozeki and T. Yamashita,
Graphs Combin. {\bf 27}, 1 (2011).

\bibitem{ZhLiWuZh10}
Z. Z. Zhang, H. X. Liu, B. Wu, and S. G. Zhou, EPL {\bf 90}, 68002
(2010).

\bibitem{ZhLiWuZo11}
Z. Z. Zhang, H. X. Liu, B. Wu, and T. Zou, Phys. Rev. E {\bf
83}, 016116 (2011).

\bibitem{Ki1847}
G. Kirchhoff, Ann. Phys. Chem. {\bf 72}, 497 (1847). 

\bibitem{ChKl78}
S. Chaiken and D. J. Kleitman,
J. Combin. Theory Ser. A {\bf 24}, 377 (1978).  

\bibitem{Bi93}
N. L. Biggs, \emph{Algebraic Graph Theory}, 2nd ed. (Cambridge
University Press, Cambridge, 1993).

\bibitem{SoHaMa05}
C. Song, S. Havlin, H. A. Makse,
Nature {\bf 433}, 392 (2005).


\bibitem{BaAl99}
A.-L. Barab\'asi and R. Albert,
       Science {\bf 286}, 509 (1999).

\bibitem{ChLu02}
F. Chung and L. Lu, Proc. Natl. Acad. Sci. U.S.A. {\bf 99}, 15879 (2002).

\bibitem{CoHa03}
R. Cohen and S. Havlin, Phys. Rev. Lett. {\bf 90}, 058701 (2003).

\bibitem{ZhZhZoChGu09}
Z. Z. Zhang, S. G. Zhou, T. Zou, L. C. Chen, and J. H. Guan, Phys.
Rev. E {\bf 79}, 031110 (2009).

\bibitem{SoHaMa06}
C. Song, S. Havlin, H. A. Makse,
Nat. Phys. {\bf 2}, 275 (2006).

\bibitem{WaSt98}
D. J. Watts and H. Strogatz,
Nature (London) {\bf 393}, 440 (1998).

\bibitem{RoHaAv07}
H. D. Rozenfeld, S. Havlin, and D. ben-Avraham, New J. Phys. {\bf
9}, 175 (2007).

\bibitem{RoAv07}
H. D. Rozenfeld and D. ben-Avraham, Phys. Rev. E {\bf 75}, 061102
(2007).

\bibitem{AlBa02}
R. Albert and A.-L. Barab\'asi,
       Rev. Mod. Phys. {\bf 74}, 47 (2002).

\bibitem{DoMe02}
S. N. Dorogvtsev and J. F. F. Mendes,
Adv. Phys. {\bf 51}, 1079 (2002).

\bibitem{Ne03}
M. E. J. Newman,
SIAM Rev. {\bf 45}, 167 (2003).


\bibitem{DoGoMe02}
S. N. Dorogovtsev, A. V. Goltsev, J. F. F. Mendes,
Phys. Rev. E {\bf 65}, 066122 (2002).

\bibitem{ZhZhCh07}
Z. Z. Zhang, S. G. Zhou, and L. C. Chen, Eur. Phys. J. B {\bf 58},
337 (2007).

\bibitem{BeOs79}
A. N. Berker and S. Ostlund, J. Phys. C {\bf 12}, 4961 (1979).

\bibitem{KaGr81}
 M. Kaufman and R. B. Griffiths,
 Phys. Rev. B {\bf 24}, 496 (1981).

\bibitem{GrKa82}
R. B. Griffiths and M. Kaufman,
Phys. Rev. B {\bf 26}, 5022 (1982).

\bibitem{HiBe06}
M. Hinczewski and A. N. Berker, Phys. Rev. E {\bf 73}, 066126
(2006).

\bibitem{BaCoDaFi09}
L. Barri\`ere, F. Comellas, C. Dalf\'o, and M. A. Fiol,
Discrete Appl. Math. {\bf 157}, 36 (2009).  

\bibitem{AnHeAnSi05}
J. S. Andrade Jr., H. J. Herrmann, R. F. S. Andrade and L. R. da Silva,
Phys. Rev. Lett. {\bf 94}, 018702 (2005).

\bibitem{ZhChZhFaGuZo08}
Z. Z. Zhang, L. C. Chen, S. G. Zhou, L. J. Fang, J. H. Guan, and T.
Zou, Phys. Rev. E {\bf 77}, 017102 (2008).

\bibitem{ZhZhXiChLiGu09}
Z. Z. Zhang,  S. G. Zhou, W. L. Xie, L. C. Chen, Y. Lin,  and J. H.
Guan, Phys. Rev. E {\bf 79}, 061113 (2009).

\bibitem{ZhGaChZhZhGu10}
Z. Z. Zhang, S. Y. Gao, L. C. Chen, S. G. Zhou, H. J. Zhang, and J.
H. Guan, J. Phys. A {\bf 43}, 395101 (2010).

\bibitem{AmScBaSt00}
L. A. N. Amaral, A. Scala, M. Barth\'el\'emy, H. E. Stanley, Proc.
Natl. Acad. Sci. U.S.A. {\bf 97}, 11149 (2000).

\bibitem{KaSu43}
E. Kasner and F. Supnick,
Proc. Natl. Acad. Sci. U.S.A. {\bf 29}, 378 (1943).


\end{references}

\end{document}